# Mining Frequent Itemsets (MFI) over Data Streams : Variable Window Size (VWS) by Context Variation Analysis (CVA) of the Streaming Transactions

V.Sidda Reddy[1], T.V.Rao[2] and A.Govardhan[3]

[1]Department of Computer and Engineering,
Sagar Institute of Technology, Chevella, Hyderabad, T.S, India
[2] Departments of Computer and Engineering,
PVP Siddhartha Institute of Technology, Vijayawada, A.P, India
[3] School of Information Technology,
Jawaharlal Nehru Technological University, Hyderabad, T.S, India

## Abstract

*The challenges with respect to mining frequent items over data streaming engaging variable window size and low memory space are addressed in this research paper. To check the varying point of context change in streaming transaction we have developed a window structure which will be in two levels and supports in fixing the window size instantly and controls the heterogeneities and assures homogeneities among transactions added to the window. To minimize the memory utilization, computational cost and improve the process scalability, this design will allow fixing the coverage or support at window level. Here in this document, an incremental mining of frequent item-sets from the window and a context variation analysis approach are being introduced. The complete technology that we are presenting in this document is named as Mining Frequent Item-sets using Variable Window Size fixed by Context Variation Analysis (MFI-VWS-CVA). There are clear boundaries among frequent and infrequent item-sets in specific item-sets. In this design we have used window size change to represent the conceptual drift in an information stream. As it were, whenever there is a problem in setting window size effectively the item-set will be infrequent. The experiments that we have executed and documented proved that the algorithm that we have designed is much efficient than that of existing.*

## 1. Introduction

Association rule is an important and well researched method for finding frequent itemsets (patterns) among set of objects in large database [1]. Frequent itemsets indicates the closeness among the items that are available in datasets. There are very good range of applications that includes creation of association rules for market-basket analysis, in text mining grading and clustering of documents, text, and pages, web mining, that allows users to disclose hidden patterns and relationships in huge datasets. The time taken to produce the data is outstripping the speed of its mining in the present upcoming applications where the information is in the form of an enormous and continuous stream [2]. To traditional static databases this is in quiet opposite, so data stream mining is considerably different from traditional data mining with respect good number of aspects. Primarily, the amount of data embedded in its lifetime in a data stream could be overwhelmingly high [3]. Along with the above it is required to create spontaneous responses by maintaining response time to queries on such information streams because of rigid pitfalls of resource [4]. The data stream mining is the area where very high level of research is taking place because of the above mentioned reasons and contemporary research area is the challenge of getting timely and appropriate association rules. Migrating from conventional data mining methods to the emerging high efficient methods those will accommodate to function on an open-ended, high speed stream of data [5]. The challenges that are inevitable to all mining technologies because of the unique and inherent qualities of a data stream are as follows [3]. Primarily the conventional technology is not applicable as it is required to create model in this method database need to be scanned multiple number of times which is not possible because of the continuous quality of the stream data. The high priority was given to the scalability of the frequent itemsets mining in approaches like association rules, classification and clustering. This requires high level of data transformation from the representation to other as the algorithm is designed in such a way, this in turn will use resources extensively resulting in high CPU overhead. Our proposal in this paper is a model known as Mining Frequent itemsets using Variable Window Size fixed by Context Variation Analysis (MFI-VWS-CVA). This model is the resource effective and measurable, as it functions with limited memory requirements and limited computational expenses. It portrays the trade-offs between computation, data representation, I/O and heuristics. Context variation analysis based dynamic window based transaction storage is being used in the proposed algorithm and also allows TIFIM [15] to regular itemsets from the concluded window.

## 2. RELATED WORK

CPS-tree a prefix-tree structure was proposed by Syed Khairuzzaman Tanbeer [6]. Dynamic tree restructuring technique was used in the CPS tree in order to manage the stream data. The basic pitfall of this model is, it reconstructs the tree for every new arrival of the item. This used to result in high memory usage and a time consuming process. Weighted Sliding Window (WSW) algorithm was initiated by Pauray S.M. Tsai [7]. In this method weight of each transaction in each window will be calculated by the algorithm proposed. Here also same memory space and time are the major concerns; this method failed using these two economically. An apriori algorithm is being for candidate generation. Hue-Fu Li [8] introduced an effective bit-sequence dependent algorithm named as MFI-TransSW (Mining Frequent Item sets with in a Transaction Sensitive Sliding window). There as three phases in MFI algorithm. As there is any increase in the window size, there will be a subsequent increase in the memory usage of MFI-TransSW. Similarly whenever there is an increase in window size, the time consumption in phase 1 and phase 2 of MFI-TransSW to process will also increase. Yo unghee Kim [9, 16] initiated an effective algorithm with normalized weight over data stream called WSFI mine (Weighted Support Frequent Item sets mining). From the database in one scan this WSFI-mine algorithm can mine all frequent item sets. HUPMS (High Utility Pattern Mining in Stream data) was recommended by Chowdhury Farhan Ahmed [10]. This is a different algorithm for sliding window based high utility pattern mining over data stream. For interactive mining only this algorithm is suitable. In the paper that was presented by Jing Guo [11], they have discussed about how to mine regular patterns across multiple data streams. Here for analysis they have considered real time news paper data. In multiple streams it is vital to identify collaborative frequent patterns and comparative frequent patterns. Prevention of misuse of sensitive data in a stream was addressed by Anushree Gowtham Ringe [12] in their work. They initiate a new technique for guarding the privacy of data stream. Fan Guidan [13] proposed a model which is conceptually similar to our model. Matrix in sliding window over data streams plays major role in this model. In order to store, this algorithms uses two 0-1 matrices and 2-itemsets, further applies relative operation of the two matrices to extract frequent itemsets.

## 3. MINING FREQUENT ITEMSETS OVER DATA STREAMS WITH CONTEXT AWARE VARIABLE SIZE WINDOWS

If streaming data is input to mining strategies such as frequent itemsets mining, the traditional approaches are not suitable, since those are mainly works by the multiple passes through entire dataset. Henceforth the mining strategies opted for streaming data considers the tuples of the streaming transactions as windows and these windows are used as input to the mining algorithms. The significant issue here in this model is fixing the window size. In the case of data streams with transitional and temporal state transactions, the transitional and temporal state identifications can be used to fix the window size. In the other cases that are not having any transitional and temporal state identities for streaming transactions, fixing window size is a big constraint to achieve quality factors such as results accuracy, process scalability. In this regard here we propose a novel context variation based dynamic window size fixing approach to mine frequent itemsets over data streams. The proposed frequent itemsets mining strategy is centric to following qualities targeted.

- The window size should be optimal and dynamic.
- The window size should fix dynamically between minimal and maximal size given as thresholds.
- The size of the window should be within the range of minimal and maximal size and should fix based on the context variation observed in input transaction from the data stream.

In regard to implementing the proposed model, the only significant constraint related to the streaming data is that the context change of the transactions should be in an order.

The minimal memory utilization and less computational cost are two main quality metrics expected from this proposal. The exploration of the proposed window fixing strategy is follows:

Let $ds$ be the DataStream, and stream transactions as horizontal partitions of the transactions, let each partition having one transaction. Let $n$ be the total count of the attributes used to form the transactions by $ds$. Let $a_{set}$ be the attributes set that contains attributes $\{a_1, a_2, \ldots\ldots a_i, a_{i+1}, \ldots\ldots a_n\}$, which are used to form the transactions. Let $\{t_1, t_2, t_3, \ldots, t_i, t_{i+1}, \ldots t_{i+m}, t_{i+(m+1)}, \ldots\ldots\}$ be the transactions streaming in the same sequence. Let $ws_{\min}$ be the minimum window size and $ws_{\max}$ be the maximal window size. Let $w_{tran}$ be the transaction window and $w_{cca}$ be the context change analysis window. Let $s(w_{cca})$ be the size of $w_{cca}$. The initial values to $ws_{\min}, ws_{\max}$ and $s(w_{cca})$ will be set during the pre-processing step.

The transactions of count $ws_{\min}$ from the given data stream $ds$ will be moved initially to $w_{tran}$, then following transactions of count $s(w_{cca})$ will be moved to $w_{cca}$. Then context variation analysis (CVA) process will be initialized. The exploration of the CVA process is follows:

The attributes involved to generate the transactions moved into $w_{tran}$ will be collected as $al(w_{tran})$, and attributes involved to form the transactions found in $w_{cca}$ will also be collected as $al(w_{cca})$. Then the similarity score of these two attribute lists $al(w_{tran}), al(w_{cca})$ will be found as follows (Eq1), which is derived from jaccard similarity measuring approach.

$$ss_{(w_{trans} \leftrightarrow w_{cca})} = \frac{al(w_{tran}) \, \mathrm{I} \, al(w_{cca})}{al(w_{trans}) \, \mathrm{U} \, al(w_{cca})} \ldots \text{(Eq1)}$$

If similarity score $ss_{(w_{tran} \leftrightarrow w_{cca})}$ is greater than the given similarity score threshold $ss_{\tau}$ then the transactions of $w_{cca}$ will be moved to $w_{tran}$ (see Eq2).

$$w_{tran} = w_{tran} \, \mathrm{U} \, w_{cca} \ldots \text{(Eq2)}$$

If size of the $w_{tran}$ is greater than or equals to $ws_{max}$ then the $w_{tran}$ will be finalized and initiates process of mining frequent itemsets from the transactions of $w_{tran}$, else the further streaming transactions of size $s(w_{cca})$ will moved to $w_{cca}$ and continues CVA process.

Once the $w_{tran}$ is finalized and mining of frequent itemsets is initiated, then $w_{tran}$ and $w_{cca}$ will be cleared and continues the process explored to prepare the window $w_{tran}$ will be continued for further transactions streaming from data stream $ds$.

The above said process continues till transactions found from data stream $ds$. The mining frequent itemsets from the finalized window will be done by using TIFIM [15], which is a tree based incremental frequent itemsets mining approach that we devised in our earlier research paper (see sec 3.3).

## 3.1 Algorithmic exploration of the Fixing Variable Window Size by Context Variation Analysis

**Inputs:**
- Data stream $ds$
- Minimal transaction window size $ws_{min}$
- Maximal transaction window size $ws_{max}$
- Similarity score threshold $ss_\tau$
- Size of the context change analysis window $s(w_{cca})$

---

1. Begin
2. For each transaction $\{t \forall t \in ds\}$ Begin
3. If $(| w_{tran} |< ws_{min})$ $w_{tran} \leftarrow t$
4. Else Begin
5. $w_{cca} \leftarrow t$
6. If $(| w_{cca} |\geq s(w_{cca}))$
7. $ss \leftarrow CVA(w_{tran}, w_{cca})$
8. if $(ss \geq ss_\tau)$ Begin
9. $w_{tran} \leftarrow (w_{tran} \cup w_{cca})$
10. If $(| w_{tran} |\geq ws_{max})$ Begin
11. Finalize window $w_{tran}$
12. Initiate $TIFIM(w_{tran})$
13. set $w_{tran} \leftarrow \phi$ // empty $w_{tran}$
14. set $w_{cca} \leftarrow \phi$ //empty $w_{cca}$
15. End of 10
16. End of 8
17. Else Begin
18. Finalize window $w_{tran}$
19. Initiate $TIFIM(w_{tran})$
20. set $w_{tran} \leftarrow \phi$ // empty $w_{tran}$
21. set $w_{tran} \leftarrow w_{cca}$ //move transactions of window $w_{cca}$ to new window $w_{tran}$

22. set $w_{cca} \leftarrow \phi$ //empty $w_{cca}$
23. End of 17
24. End of 4
25. End of 2
26. End of 1

## 3.2 Algorithmic exploration of the Context Variation Analysis

1. $CVA(w_{tran}, w_{cca})$ Begin
2. Set $fs_{tran} \leftarrow \phi$ // initiate field set $fs_{tran}$ of transaction window $w_{tran}$ empty
3. Foreach transaction $\{t \forall t \in w_{tran}\}$ Begin
4. $fs_{tran} \leftarrow fs_{tran} \cup t$
5. End of 3
6. Set $fs_{cca} \leftarrow \phi$ // initiate field set $fs_{cca}$ of transaction window $w_{cca}$ empty
7. Foreach transaction $\{t \forall t \in w_{cca}\}$ Begin
8. $fs_{cca} \leftarrow fs_{cca} \cup t$
9. End of 7
10. $ss = \dfrac{fs_{tran} \cap fs_{cca}}{fs_{tran} \cup fs_{cca}}$ //measuring similarity score of $w_{tran}$ and $w_{cca}$
11. Return $ss$
12. End of 1

## 3.3 Tree (Bush) Based Incremental Frequent Itemsets Mining (TIFIM)

### 3.3.1 Finding Frequent Itemsets

The primary representation of the transactions of data stream $ds$ is as described above. An asynchronous parallel process runs to find frequent itemsets in incremental manner.

A bush represents itemsets with two attribute pair such that these attributes belongs to $fs_{tran}$ and transactions contain that pair. The coverage is to measure the frequency of the itemsets can be considered and set in the context of window $w_{tran}$ size. The coverage of two attribute itemsets can be the count of number of Childs in a bush represented by each pair of attributes.

An asynchronous parallel process called frequent itemsets finder (FIF) performs as follow:
Initially picks the bushes with coverage more than given coverage threshold $cov$.
Prepare new bushes from each two bushes by union the roots and intersects the Childs, and retains it if new bush coverage is greater or equal to $cov$ else discards.
This continues until no new bush formed.

### 3.3.2  The pruning process

A bush $b_i$ said to be sub-bush to bush $b_j$ if $r_{b_i} \subseteq r_{b_j}$ and $cov_{(b_i)} \leq cov_{(b_j)}$. Since sub-bush $b_i$ represented by $b_j$, then bush $b_i$ can be pruned from the bush-set $B$.

## 3.4 Find frequent items

At an event of time, frequent itemsets can be found as follows
The roots of the bushes with coverage more than given $cov$ can be claimed as frequent itemsets.
A bush '$b_i$' coverage can be find as follows

If a bush $b_j$ found to be such that $b_i \subseteq b_j$ and coverage value of $b_j$ is higher than any other bush $b_k$ such that $b_i \subseteq b_k$, then the coverage of $b_i$ said to be $\mathrm{cov}_{(b_j)} + \mathrm{cov}_{(b_i)}$.

### 3.4.1 An algorithmic representation of the caching processes

**Input**: At an event of time a window $w_i$ with transaction $t_i$ received

For each transaction $t_i$:

$$\{(a_1, a_2, a_3, .........a_i)\forall$$

Let set of attributes $(a_1, a_2, a_3, .........a_i) \in t_i \wedge$

$$(a_1, a_2, a_3, .........a_i) \subseteq A_{set}\}$$

For each pair of attributes $\{(a_m, a_n)\forall (a_m, a_n) \in t_i\}$, if found a bush $\{b_i \exists (a_m, a_n)\ as\ root\}$ then add transaction $t_i$ as node to bush $b_i$, else prepare a bush such that $\{b_i \exists (a_m, a_n)\ as\ roo\,t \wedge t_i\ as\ node\}$

### 3.4.2 An algorithmic approach of FIF

The bush set $B$ prepared by caching process is said to be input to FIF

For each bush $\{b_i \forall b_i \in B\}$ perform the following:

For each bush $b_{i+c} \exists (c := 1, 2, 3.......n) \wedge b_{i+c} \in B)$

Forms a bush $\{b_{(iUi+c)} \exists b_{(iUi+c)} \notin B\}$ by Union the roots of the '$b_i$' and '$b_{i+c}$' ($r_{(b_i)}$ U $r_{(b_{i+c})}$) and intersects nodes of $b_i$ and $b_{i+c}$ ($ts_{(b_i)}$ I $ts_{(b_{i+c})}$).

## 4. EMPIRICAL ANALYSIS OF THE FIXING VARIABLE WINDOW SIZE BY CONTEXT VARIATION ANALYSIS

### 4.1 Dataset characteristics

Multiple sets of data streamed to perform the experiments, and the characteristics of these streaming data are as follows:

- To achieve the sparseness in streaming transactions, the range of fields considered as 75,100, 125 and 150, the max transaction length set in the range of 12 to 18, the min transaction range set to 5 and the total number of transactions has taken in the range of 1000 to 10000.
- To achieve the denseness in streaming transactions, the range of fields considered as 20, 30, 40 and 50, the max transaction length set in the range of 10 to 15, the min transaction range set to 5 and the total number of transactions has taken in the range of 1000 to 10000.

### 4.2 Experimental results

We compare our algorithm with frequent itemsets mining model for data streams devised in [13], which is a matrix based frequent itemsets mining (MFIM) algorithm for data streams. The implementation of our MFI-VWS-CVA and model MFIM done by using java 7 and set of flat files as streaming data sources. The streaming environment is emulated using java RMI and parallel process involved in proposed MFI-VWS-CVA is achieved by using java multi threading concept. The three parameters of each synthetic dataset are the total number of transactions, the average length, and divergence count of items, respectively. Each transaction of a dataset is scanned only once in our experiments to simulate the environment of data streams. In regard to measure the computational cost and scalability, the algorithms run under divergent coverage values in the range of 10% to 90%.

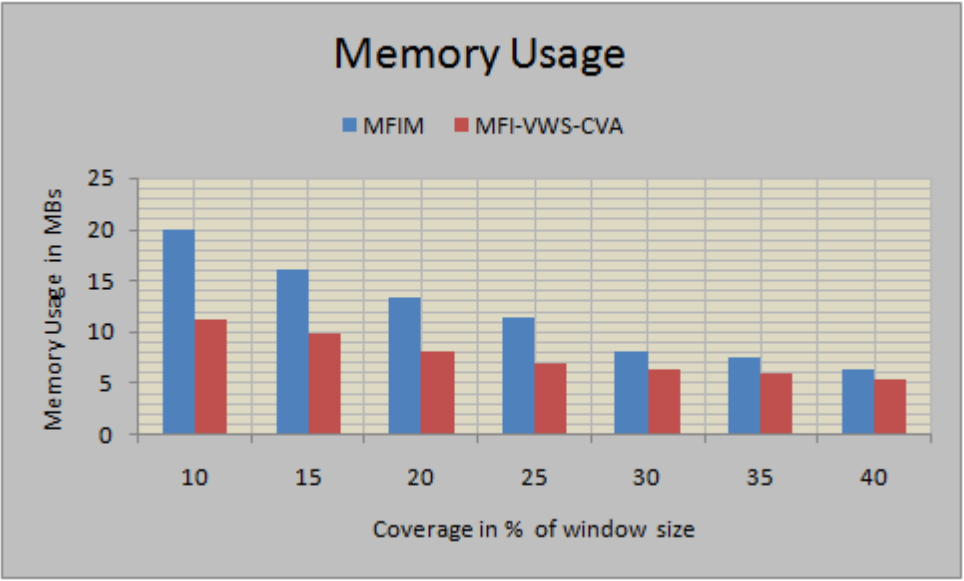

Figure 1: NFI-VWS-CVA advantage over MFIM in Memory usage.

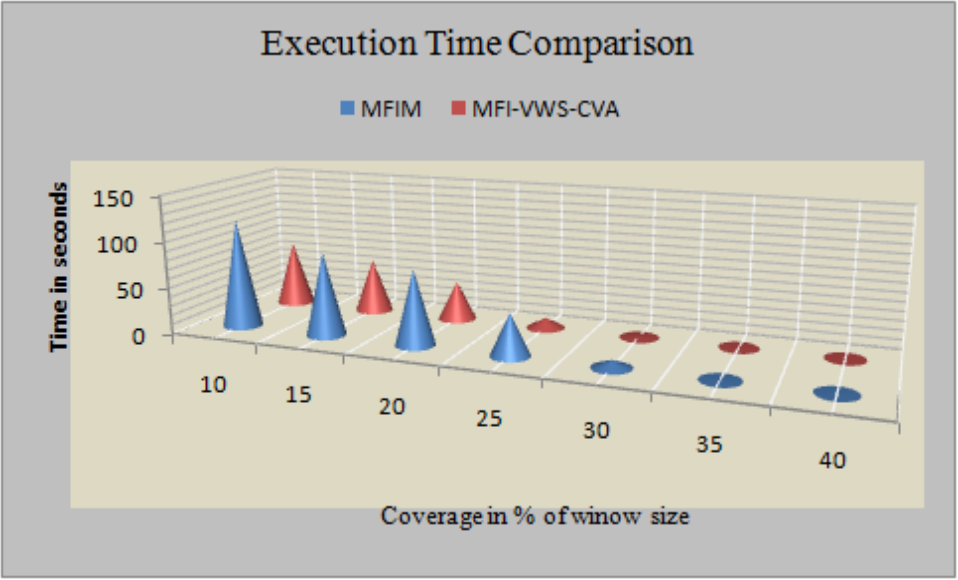

Figure 2: MFI-VWS-CVA advantage over MFIM in terms of execution time.

Figure 1 and 2 shows the comparison of the Memory usage, execution time under divergent coverage values range given from 10% to 40% respectively. The Figure 3 explores the scalability of MFI-VWS-CVA over MFIM under divergent streaming data sizes respectively, In Figure 1 and 2, the horizontal axis is the coverage given and the vertical axis is the memory in unit of mega bytes and time in unit of seconds respectively. In Figure 3, the horizontal axis is the streaming data size given in unit of transactions and the vertical axis is the execution time in unit of seconds and percentage of elapsed time in unit of seconds respectively. As the coverage value decreases the average increment in memory usage for matrix based FIM and MFI-VWS-CVA are 2.29 and 0.7 respectively (see Figure 1) and average execution time increment for matrix based FIM and MFI-VWS-CVA are 83.2 and 27.9 respectively (see Figure 2). The results obtained here clearly indicating that the performance of MFI-VWS-CVA is miles ahead than the matrix based FIM. The performance of MFI-VWS-CVA is scalable as matrix based FIM is taking average of 14.16% elapsed time under uniform increment of streaming data size with 1500 transactions.

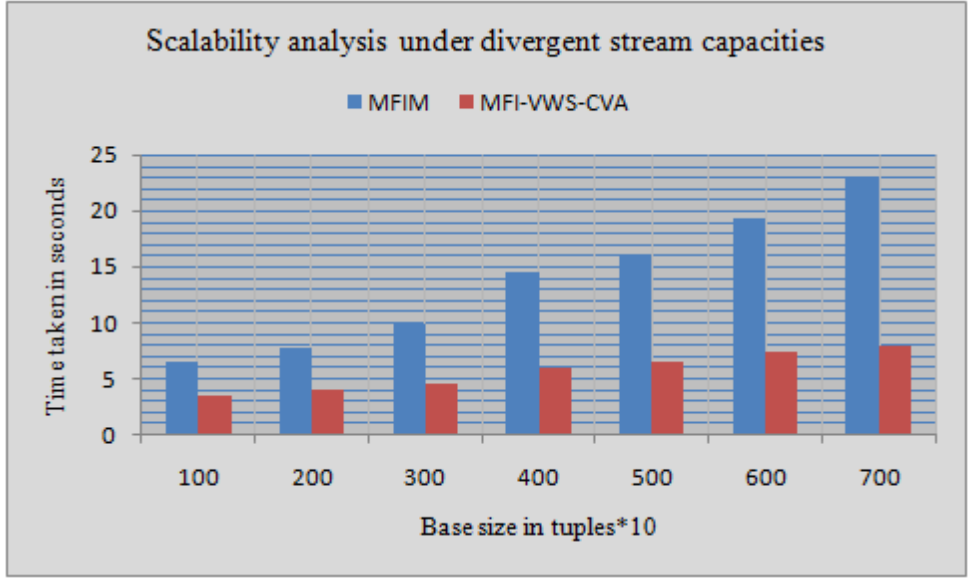

Figure 3: MFI-VWS-CVA advantage over MFIM about Scalability under divergent streaming data size.

## 5. CONCLUSION

We explored a novel approach for mining the frequent itemsets from a data stream. We have implemented an efficient tree based incremental frequent itemsets mining model [15] in our earlier research paper, further here we developed an approach for mining frequent itemsets using variable window size by context variation analysis (MFI-VWS-CVA) over data streams. Due to the factor of fixing window size dynamically by concept variation analysis, the said model is identified as optimal and scalable. A parallel process that determines frequent itemsets from the concept of cached bush structures, which is our earlier proposal [15], performs frequent itemsets mining over data streams. We extended this incremental frequent itemset mining algorithm by introducing windowing the streaming transaction with variable window size technique in regard to achieve efficient memory usage and execution time. The experiment results confirm that the MFI-VWS-CVA is scalable under divergent streaming data size and coverage values. In future this model can be extended to perform utility based frequent itemset mining over data streams.

## ACKNOWLEDGEMENT

The authors would like to thank the anonymous reviewers for their valuable comments. We also thank the authors of all references for helping us setup the paper.